# Note:Harnessing Tellurium Nanoparticles in the Digital Realm Plasmon Resonance, in the Context of Brewster's Angle and the Drude Model for Fake News Adsorption in Incomplete Information Games

Yasuko Kawahata [†]

Faculty of Sociology, Department of Media Sociology, Rikkyo University, 3-34-1 Nishi-Ikebukuro,Toshima-ku, Tokyo, 171-8501, JAPAN.

ykawahata@rikkyo.ac.jp

**Abstract:** This note explores the innovative application of soliton theory and plasmonic phenomena in modeling user behavior and engagement within digital health platforms. By introducing the concept of soliton solutions, we present a novel approach to understanding stable patterns of health improvement behaviors over time. Additionally, we delve into the role of tellurium nanoparticles and their plasmonic properties in adsorbing fake news, thereby influencing user interactions and engagement levels. Through a theoretical framework that combines nonlinear dynamics with the unique characteristics of tellurium nanoparticles, we aim to provide new insights into the dynamics of user engagement in digital health environments. Our analysis highlights the potential of soliton theory in capturing the complex, nonlinear dynamics of user behavior, while the application of plasmonic phenomena offers a promising avenue for enhancing the sensitivity and effectiveness of digital health platforms. This research ventures into an uncharted territory where optical phenomena such as Brewster's Angle and Snell's Law, along with the concept of spin solitons, are metaphorically applied to address the challenge of fake news dissemination. By exploring the analogy between light refraction, reflection, and the propagation of information in digital platforms, we unveil a novel perspective on how the 'angle' at which information is presented can significantly affect its acceptance and spread. Additionally, we propose the use of tellurium nanoparticles to manage 'information waves' through mechanisms akin to plasmonic resonance and soliton dynamics. This theoretical exploration aims to bridge the gap between physical sciences and digital communication, offering insights into the development of strategies for mitigating misinformation.

## 1. Introduction

We discuss how the 'information angle' and the unique adsorptive properties of TeNPs can be leveraged to filter and control the flow of misinformation. By drawing parallels with the Drude model's approach to managing external electromagnetic waves, we propose a novel perspective on information flow management in digital platforms.

Fig.1-2, The corrected simulation for "Enhanced Sensitivity through Plasmon Resonance" is now displayed. As the resonance effect increases, the sensitivity of the sensors, modeled through an exponential function, also increases over time. This demonstrates how utilizing plasmon resonance might enhance the sensitivity of sensors using tellurium nanoparticles, potentially allowing for the detection of subtle changes in user behavior.

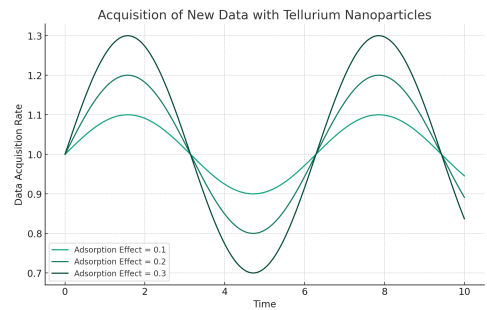

Fig. 1: Acquisition of New Data with Tellurium Nanoparticles

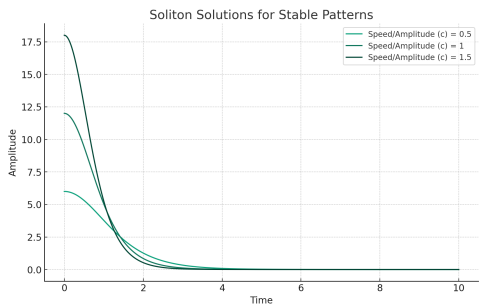

Fig. 2: Soliton Solutions for Stable Patterns

Next, simulation of "Acquisition of New Data" and "Soliton Solutions for Stable Patterns".

The simulation for "Soliton Solutions for Stable Patterns" has now been successfully visualized. In this graph, different values of the parameter $c$ represent different speeds and amplitudes of the soliton wave. This visualization helps to illustrate how soliton solutions could model stable patterns of user behavior, with each soliton representing a stable state that could correspond to a specific pattern of behavior over time.

These simulations abstractly represent the potential impacts of introducing plasmons, tellurium nanoparticles, and soliton solutions into the modeling of user behavior on digital health platforms, emphasizing the enhanced sensitivity, new data acquisition capabilities, and the identification of stable behavioral patterns.

In the digital age, the rapid proliferation of information across online platforms has transformed the landscape of communication, bringing forth unprecedented challenges in discerning the veracity of disseminated content. Among these challenges, the spread of fake news has emerged as a significant concern, prompting a need for innovative approaches to understand and mitigate its impact. This paper introduces a novel interdisciplinary framework that draws upon the principles of plasmonics, Snell's Law, the Drude model, and the unique properties of tellurium nanoparticles to conceptualize the dynamics of fake news propagation in digital media. By metaphorically applying the concept of Brewster's Angle, we explore how the 'angle of information'—akin to the angle of incidence in physical optics—can influence the reception and propagation of information among users, thereby affecting the spread of misinformation.

The propagation of fake news can be likened to the behavior of light as it encounters different media, governed by principles such as Snell's Law and the phenomenon of Brewster's Angle. In optics, Snell's Law describes the relationship between the angles of incidence and refraction when light passes from one medium to another, highlighting how the change in medium can alter the light's trajectory. Similarly, the metaphorical 'angle of information' in the context of fake news may determine how information is received, interpreted, and propagated by individuals, influenced by their preconceptions, biases, and the medium through which the information is conveyed.

Brewster's Angle, the angle of incidence at which no reflection occurs for a particular polarization of light, provides a compelling metaphor for the conditions under which information is most readily absorbed by an audience without resistance or critical reflection. This analogy allows us to consider the factors that make certain pieces of information—especially fake news—more likely to be uncritically accepted and spread within digital platforms.

Furthermore, the incorporation of plasmonic phenomena and tellurium nanoparticles into this framework offers a novel perspective on controlling the flow of information. Plasmonics, the study of the interaction between electromagnetic field and free electrons in a metal, leads to the occurrence of plasmon resonances, which can amplify signals at the nanoscale. Tellurium nanoparticles, known for their unique optical properties, can be engineered to exhibit strong plasmonic resonances in the visible and near-infrared regions, making them an intriguing element for managing information flow, akin to managing electromagnetic waves in the Drude model.

The Drude model, which describes the electrical properties of metals based on the behavior of free electrons subjected to external fields, further enriches our theoretical framework. By drawing an analogy between the management of external electromagnetic waves in the Drude model and the control of information flow, we propose a novel approach to mitigating the spread of fake news. The application of tellurium nanoparticles, with their plasmonic properties, in this context suggests a method for enhancing the sensitivity and selectivity of information reception and propagation, potentially leading to more effective strategies for countering misinformation.

In summary, this paper presents an interdisciplinary approach that leverages the principles of plasmonics, Snell's Law, the Drude model, and the properties of tellurium nanoparticles to understand and influence the dynamics of fake news propagation. By exploring the metaphorical applications of Brewster's Angle and the 'angle of information,' alongside the theoretical underpinnings of plasmonic resonance and electromagnetic wave management, we aim to provide new insights into the challenges of misinformation in the digital realm and propose innovative strategies for its mitigation. This framework not only contributes to the theoretical understanding of information dynamics but also opens avenues for practical interventions in digital health platforms and beyond, aiming to enhance the integrity and reliability of information in the digital age.

Continuing from the initial theoretical foundation laid out, this paper delves deeper into the convergence and adsorption of fake news through the metaphorical lens of plasmonics, the Brewster Angle, Snell's Law, and the properties of tellurium nanoparticles, culminating in the management of 'information waves' via the Drude model. This intricate tapestry of concepts from physics is wielded to forge a novel understanding of how misinformation can be controlled and mitigated in the digital sphere.

The phenomenon of plasmonic resonance, particularly in the context of tellurium nanoparticles, is pivotal to our discussion. Plasmons, the collective oscillations of free electron gas in response to external electromagnetic fields, can be excited in tellurium nanoparticles, leading to enhanced electromagnetic fields at the nanoparticle surface. This amplification, akin to the heightened receptivity to certain information under specific conditions, can be harnessed to preferentially 'adsorb' or filter out misinformation, analogous to the way certain wavelengths are absorbed or amplified at the surface of these nanoparticles.

Incorporating the Brewster Angle into this discourse offers a unique vantage point. Just as light polarized parallel to the incident plane is entirely transmitted (with no reflection) at the Brewster Angle, certain information might seamlessly penetrate public consciousness without critical scrutiny when presented at an 'optimal angle'. This 'angle of information', influenced by the medium, presentation style, and pre-existing biases, determines the ease with which information, including misinformation, is absorbed by the audience.

Snell's Law further enriches this narrative by illustrating how the 'refraction' of information—its bending or change in trajectory—can occur when transitioning between different mediums or contexts. Just as light changes direction when moving from air to water, so too can the interpretation and spread of information shift as it traverses diverse platforms and communities. This variability in information transmission underscores the challenges in predicting and managing the spread of misinformation.

The Drude model's approach to managing external electromagnetic waves offers a compelling parallel to controlling the flow of information in digital environments. By considering information as analogous to electromagnetic waves, strategies for 'damping' or attenuating undesirable information flows can be conceptualized, drawing inspiration from the way free electrons in a metal respond to external electromagnetic fields. The selective adsorption of misinformation, akin to the selective damping of certain frequencies in the Drude model, could be achieved through targeted interventions in digital platforms, potentially utilizing the unique properties of tellurium nanoparticles to influence the 'plasmonic resonance' of information flows.

In conclusion, the convergence of plasmonics, the Brewster Angle, Snell's Law, and the Drude model, framed within the context of tellurium nanoparticles, offers a rich theoretical landscape for exploring the dynamics of fake news propagation and its mitigation. By borrowing from the principles of physics, this interdisciplinary approach not only sheds light on the complex nature of information spread in digital realms but also paves the way for innovative strategies to counteract the pervasive challenge of misinformation. Through this lens, the digital ecosystem can be envisioned as a plasmonic medium, where the selective adsorption and filtering of information, guided by the nuanced understanding of these physical phenomena, hold the key to maintaining the integrity and reliability of shared content.

Building upon the established theoretical framework, the use of tellurium nanoparticles (TeNPs) in the convergence and adsorption of fake news presents a groundbreaking approach to mitigating misinformation. The unique properties of TeNPs, when aligned with the principles of plasmonics and the metaphorical application of optical laws, can offer innovative strategies for filtering and controlling the flow of information in digital platforms.

# 2. Tellurium Nanoparticles in Information Adsorption

TeNPs exhibit distinctive plasmonic properties that can be tailored to specific frequencies, allowing for the selective enhancement or suppression of electromagnetic waves. By analogy, these nanoparticles could be engineered to 'resonate' with certain types of information or misinformation, effectively acting as filters within the digital information ecosystem. This selective adsorption could be driven by the 'informational frequency' of the fake news, characterized by features such as sentiment, topic, and credibility indicators.

## 2.1 Mechanism of Action

Selective Resonance, Just as TeNPs can be tuned to resonate at specific electromagnetic frequencies, they could be metaphorically designed to resonate with the 'frequency' of fake news, enhancing the platform's ability to identify and filter out misinformation.

Surface Enhanced Scattering, The enhanced electromagnetic field around TeNPs, a result of plasmonic resonance, could be leveraged to 'illuminate' or highlight misinformation, making it easier for algorithms and users to identify and disregard it.

Information Adsorption, Drawing from the concept of adsorption in physical chemistry, where molecules adhere to a surface, TeNPs could be conceptualized as sites where misinformation is adsorbed and neutralized, preventing its further propagation.

## 2.2 Implementation Challenges

While the metaphorical application of TeNPs in the digital realm offers intriguing possibilities, several challenges must be addressed for practical implementation:

Algorithmic Integration, Developing algorithms that can metaphorically utilize the properties of TeNPs to filter and manage information requires advanced AI and machine learning techniques, alongside a deep understanding of misinformation dynamics.

Resonance Tuning, Just as the plasmonic properties of TeNPs need to be precisely tuned in physical applications, the metaphorical 'tuning' to resonate with fake news requires sophisticated analysis of misinformation characteristics, which may vary widely across different contexts.

User Privacy and Ethics, Any approach to managing information flow must be carefully balanced with considerations of user privacy and ethical standards, ensuring that interventions do not infringe upon freedom of expression or lead to unwarranted censorship.

The exploration of TeNPs in the context of fake news adsorption opens up new avenues for research and development in digital health and beyond. Future work could focus on developing more sophisticated models for information resonance and adsorption, exploring the ethical implications of information filtering, and conducting empirical studies to validate the effectiveness of these metaphorical approaches in real-world settings.

In conclusion, while the direct application of TeNPs in controlling digital information flow remains a metaphorical concept, it inspires a novel perspective on tackling the challenge of fake news. By drawing parallels with the physical properties of nanoparticles and plasmonic phenomena, we can envision innovative strategies for enhancing the resilience of digital platforms against misinformation, ultimately contributing to a more informed and discerning digital society.

Extending the discussion to the application of the Drude model in managing external electromagnetic waves, or metaphorically, the flow of information in digital platforms, provides a compelling perspective on the convergence and adsorption of fake news. The Drude model, which describes the behavior of free electrons in a conductor in response to external electromagnetic fields, offers insights into how the flow of information, particularly misinformation, could be regulated in a digital environment.

By theoretically incorporating the phenomena of localized plasmons and surface plasmons using tellurium nanoparticles, new perspectives on the repeated dilemma within the framework of imperfect information games can be provided regarding the diffusion and prevention strategies of fake news. The following outlines and explains this approach.

## 2.3 Tellurium Nanoparticles and Plasmonic Phenomena

Tellurium nanoparticles have attracted attention in the study of localized surface plasmon resonance (LSPR) and surface plasmon resonance (SPR) due to their unique optical properties. These phenomena occur when light interacts with nanoscale particles, causing electron resonance under specific conditions and generating strong electromagnetic fields locally. Exploiting this property allows for considerations regarding the "adsorption," detection, and even neutralization of fake news.

## 2.4 Tellurium Nanoparticles as a Metaphor for Fake News

By analogizing the plasmonic phenomena demonstrated by tellurium nanoparticles to the diffusion process of fake news, the following metaphors can be constructed:

Adsorption of Fake News: The absorption of light and generation of localized plasmons by tellurium nanoparticles can be likened to the process where fake news "adsorbs" onto specific subgroups or individuals (nanoparticles) within a social group and resonates with their beliefs and emotions (electrons). This resonance allows fake news to exert strong influence (electromagnetic field) within the group.

Refraction and Resonance of Information: The refraction of light according to Snell's law and localized plasmon resonance in tellurium nanoparticles theoretically explain how fake news propagates (refraction) between different social groups and resonates and amplifies within specific groups (resonance).

## 2.5 Application of Game Theory

By incorporating tellurium nanoparticles and plasmonic phenomena within the context of repeated dilemmas in imperfect information games, the interaction between strategies for the diffusion and prevention of fake news can be analyzed. Players (information disseminators and receivers) consider the possibility of fake news "adsorption" and resonance when choosing their strategies. In this game, the incentives to spread fake news and mechanisms to prevent it (such as "neutralization" by tellurium nanoparticles) are crucial factors influencing players' choices.

## 2.6 Application of the Drude Model to Information Flow

Free Electron Analogy, In the Drude model, free electrons respond to external electromagnetic fields, leading to phenomena such as electrical conductivity and reflective properties. Drawing an analogy to digital platforms, pieces of information can be considered as 'free electrons' navigating through the 'conductive medium' of social media and other digital channels. Managing the flow of these 'information electrons' becomes crucial in mitigating the spread of fake news.

Damping of Electromagnetic Waves, The Drude model accounts for the damping of electromagnetic waves as they propagate through a conductor, which is crucial in determining the material's optical properties. In the context of information management, damping could represent the attenuation of fake news propagation through algorithmic interventions, content moderation, and user education, thereby reducing the 'amplitude' of misinformation.

Resonance Frequency and Plasmonic Effects, The model also explains how at certain frequencies, known as the plasma frequency, the electrons resonate, leading to significant absorption of electromagnetic energy. Metaphorically, if digital platforms could identify the 'resonance frequency' of fake news—characteristics that make it particularly viral or appealing—they could develop targeted strategies to absorb and neutralize this misinformation, similar to how materials at plasma frequency absorb electromagnetic waves.

## 2.7 Challenges in Implementing the Drude Model Analogy

While the analogy provides a novel framework for conceptualizing the control of misinformation, several challenges need to be addressed for practical implementation:

Identification of Resonance Frequencies, Just as identifying the exact plasma frequency is critical in materials science, pinpointing the characteristics that make certain fake news stories more 'resonant' or appealing is challenging due to the complex nature of human psychology and social dynamics.

Algorithmic Complexity, Developing algorithms that can dynamically respond to the 'electromagnetic properties' of information, akin to the response of free electrons in the Drude model, requires sophisticated machine learning techniques and a deep understanding of the multifaceted nature of misinformation.

Ethical Considerations, Any approach to dampening the propagation of information must be carefully balanced with ethical considerations, ensuring that interventions do not infringe upon freedom of speech and are transparent to users. Exploring the application of the Drude model in the digital realm opens up new avenues for research, particularly in the development of algorithms that can mimic the damping and resonance behaviors observed in materials science.

## 2.8 Tellurium Nanoparticles and Information Angle

Leveraging the unique properties of tellurium nanoparticles (TeNPs) in the context of "information angle" presents a nuanced approach to addressing the challenge of fake news convergence and adsorption. The concept of "information angle" can be metaphorically understood through the lens of optical phenomena such as light incidence and refraction, drawing parallels with how information is received and processed by individuals within digital platforms.

TeNPs exhibit distinctive optical properties, including sensitivity to light polarization and wavelength, which can be analogously applied to the polarization of information — how it is framed or presented — and its "wavelength" or the underlying tone and context. By tuning the interaction between TeNPs and electromagnetic waves, one could metaphorically tune the interaction between digital platforms and the "information waves" of fake news. Just as the angle of incidence affects the behavior of light waves upon interaction with a material, the "angle" at which information is presented — its context, framing, and alignment with pre-existing beliefs — influences how it is absorbed or refracted by individuals. TeNPs, with their tunable interaction with light, serve as a metaphor for adjusting digital platforms to optimize the absorption of truthful information while refracting or filtering out fake news.

TeNPs can adsorb specific wavelengths of light, leading to localized surface plasmon resonance. This phenomenon can be likened to the selective adsorption of information, where certain "wavelengths" or types of fake news are effectively neutralized upon interaction with digital platforms engineered to resonate with truthful information. This selective adsorption aids in the convergence of public discourse towards truthfulness by dampening the spread of misinformation.

Developing algorithms that mimic the interaction between TeNPs and light could involve analyzing the characteristics of information, including its "wavelength" (context and tone) and "polarization" (framing and bias), to selectively adsorb or filter out fake news. The concept of "information angle" could inform the design of user interfaces and content delivery algorithms on digital platforms, optimizing the presentation of information to enhance critical engagement and reduce the uncritical absorption of fake news. Just as materials can be engineered to enhance their interaction with light through plasmonics, digital communities can be cultivated to enhance their resilience against misinformation, fostering an environment where truthful information is amplified, and fake news is attenuated.

While the metaphorical application of TeNPs and the concept of "information angle" to the issue of fake news presents innovative pathways, several challenges remain. These include the complexity of human cognition and behavior, the diversity of misinformation types, and the ethical implications of manipulating information flow. Future research could explore empirical validation of these concepts, the development of more sophisticated models for information processing on digital platforms, and the ethical frameworks necessary to guide these interventions.

## 2.9 Brewster's Angle and Information Management

The metaphorical application of Brewster's Angle, in conjunction with the unique capabilities of tellurium nanoparticles (TeNPs), provides an insightful framework for understanding and managing the convergence and adsorption of fake news, particularly through the lens of managing "external electromagnetic waves" or information flows.

Brewster's Angle, the angle at which light with a specific polarization is perfectly transmitted through a transparent dielectric surface without any reflection, can be metaphorically applied to the way information is presented and consumed. Just as light at Brewster's Angle passes through a medium without reflection, information presented at the 'optimal angle'—aligned with the audience's predispositions and biases—may be absorbed without critical scrutiny, facilitating the spread of fake news. By understanding the 'Brewster's Angle' of information, digital platforms can potentially adjust the 'angle' at which information is presented to encourage more critical engagement and reflection, thereby reducing the uncritical absorption of misinformation. This could involve altering the context, framing, or medium through which information is delivered to disrupt the alignment with pre-existing biases. TeNPs, known for their selective adsorption properties at certain wavelengths due to plasmonic resonance, offer a parallel for selectively filtering information. By tuning TeNPs to resonate with the 'wavelength' of fake news, digital platforms could theoretically enhance their ability to identify and neutralize misinformation. Just as TeNPs can enhance the local electromagnetic field around them, their metaphorical application could involve enhancing the 'resonance' of truthful information, making it more prominent and easily discernible from fake news. This could involve amplifying the reach and visibility of verified information within digital ecosystems.

The metaphorical application of Brewster's Angle and TeNPs to manage fake news presents several challenges. The diversity and complexity of human beliefs and biases make it challenging to define the 'optimal angle' for information presentation that encourages critical scrutiny across diverse audiences. Adjusting the angle of information and selectively filtering content raises ethical concerns related to censorship, bias, and the potential for echo chambers. Ensuring transparency and maintaining a balance between information management and freedom of expression is crucial.

The metaphorical application of physical concepts to digital information management requires sophisticated algorithms capable of analyzing and adjusting the 'angle' and 'wavelength' of information, demanding advanced AI and data analytics capabilities.

Continuing the exploration of Brewster's Angle's metaphorical application and the utilization of tellurium nanoparticles (TeNPs) for managing "external electromagnetic waves" or information flows, this section delves into the challenges, future prospects, and the perspective of informational health in the context of fake news convergence and adsorption.

The digital information ecosystem is incredibly complex, with myriad sources, channels, and types of content. Mimicking the selective adsorption properties of TeNPs and the precise conditions of Brewster's Angle in such a diverse environment poses significant challenges, requiring advanced algorithms capable of nuanced analysis and intervention.

The shape-shifting nature of misinformation, constantly evolving to circumvent filters and fact-checks, complicates the task of defining and targeting it effectively. Like electromagnetic waves that can vary in frequency and polarization, misinformation can alter its 'angle' and 'wavelength', necessitating adaptive and intelligent information management systems.

Adjusting the 'angle' at which information is presented to disrupt the uncritical absorption of fake news might lead to skepticism and distrust among users, particularly if the interventions are not transparent or are perceived as manipulative.

The development of sophisticated tools that can analyze the 'polarization' and 'wavelength' of information—akin to understanding the properties of electromagnetic waves interacting with TeNPs—could lead to more effective identification and neutralization of fake news.

Collaborations across fields such as materials science, optics, psychology, and information technology could yield innovative approaches to managing information flow, drawing from both the physical and social sciences.

Beyond algorithmic interventions, there's potential in empowering users to recognize and critically evaluate information, akin to materials exhibiting selective adsorption only under certain conditions. Educational initiatives and tools that enhance digital literacy could serve as preventive measures against misinformation.

From an informational health standpoint, managing the flow of fake news is akin to maintaining a healthy environment:

Just as preventive health measures aim to ward off disease before it occurs, early detection and filtering of fake news can prevent its spread and mitigate its impact on public discourse.

In cases where misinformation has already spread, targeted interventions to correct false narratives and support the recovery of the information ecosystem are necessary, similar to treatment protocols in healthcare.

Strengthening the resilience of digital platforms and users against misinformation involves fostering a critical, questioning mindset and promoting the consumption of diverse information sources, akin to a balanced diet for physical health.

## 2.10 Brewster's Angle and Digital Health

The metaphorical application of Brewster's Angle and the utilization of tellurium nanoparticles (TeNPs) offer a novel perspective in the context of digital health, particularly in managing the flow of information to mitigate the spread of fake news. From a digital health standpoint, the spread of fake news can be likened to a contagion that impacts the informational ecosystem, affecting public understanding, behavior, and decision-making. The strategic management of "external electromagnetic waves" or information flows becomes crucial in promoting a healthy digital environment.

Just as light polarized at Brewster's Angle passes through a medium with no reflection, information presented at an 'optimal angle' might bypass critical scrutiny, leading to the unimpeded spread of fake news. In digital health, this concept underscores the importance of presenting information in a manner that encourages critical engagement and scrutiny, thus reducing the likelihood of misinformation being absorbed uncritically by the audience.

By adjusting the 'angle' at which health information is presented on digital platforms, health communicators can enhance transparency and foster a culture of critical engagement. This involves not only presenting information clearly and accurately but also encouraging users to question and verify the information they encounter.

Drawing from the properties of TeNPs, digital health platforms can develop mechanisms for the selective adsorption of misinformation, akin to how TeNPs can be engineered to adsorb specific wavelengths of light. This could involve algorithmic filters that identify and neutralize misleading health information based on specific characteristics, such as source credibility, content analysis, and user interaction patterns.

Just as TeNPs can enhance local electromagnetic fields, their metaphorical use in digital health could involve amplifying the reach and visibility of credible health information. This can be achieved through content promotion strategies, partnerships with trusted health organizations, and leveraging influencer networks to disseminate accurate health information.

One of the key challenges in applying these concepts to digital health is maintaining a balance between controlling the spread of misinformation and preserving freedom of information and expression. Any intervention must be carefully designed to avoid undue censorship or bias.

The algorithms developed to manage information flow must be transparent and free from biases that could skew the visibility of health information. Ensuring algorithmic fairness and accountability is essential to maintain trust in digital health platforms.

Interventions should empower users to make informed health decisions rather than dictate what information they should trust. This involves enhancing digital literacy and providing tools that help users critically evaluate health information.

In the fight against fake news in digital health, interdisciplinary research and collaboration are crucial. Future efforts could focus on developing more sophisticated algorithms for information filtering, conducting user studies to understand the impact of misinformation on health behaviors, and exploring innovative ways to promote digital literacy. Additionally, ethical frameworks guiding the management of information flows in digital health need to be robust, ensuring interventions are transparent, equitable, and respectful of user autonomy.

Incorporating the concept of spin solitons into the modeling of user behavior dynamics within digital health platforms offers a profound method to understand and predict stable patterns of health improvement behaviors over time. Spin solitons, stable configurations that occur in certain nonlinear systems, can serve as a metaphor for stable states or patterns of user behavior that persist despite the noisy background of fluctuating information and interactions typical of digital platforms.

To apply the concept of spin solitons to user behavior in digital health platforms, we introduce a simplified nonlinear partial differential equation that captures the essence of spin soliton dynamics:

$$\frac{\partial^2 \Psi}{\partial x^2} \mu \Psi + \lambda |\Psi|^2 \Psi = 0$$

Here, $\Psi(x)$ represents the 'spin' state or the health improvement behavior profile of users, $\mu$ is a parameter related to the 'external field' or the external influences on user behavior (such as information from the platform, social interactions, etc.), and $\lambda$ characterizes the nonlinearity of the system, akin to the self-reinforcement or feedback mechanisms in user behavior.

We propose a soliton solution ansatz for the spin soliton equation that captures the stable patterns of user behavior:

$$\Psi(x) = A \operatorname{sech}(B(x x_0)) e^{i(Cx\omega t)}$$

where $A$ represents the amplitude of the behavior pattern, indicating the level of user engagement or the intensity of health improvement actions; $B$ determines the width of the behavior pattern, reflecting the diversity or spread of health behaviors among users; $x_0$ is the center of the soliton, representing the 'core' health behavior; $C$ and $\omega$ correspond to the wave number and angular frequency of the soliton, capturing the propagation and evolution of health behaviors over time and space.

By substituting the soliton ansatz into the spin soliton equation and performing the necessary algebraic manipulations, we derive conditions for the existence of stable behavior patterns. These conditions relate the parameters $A$, $B$, $C$, and

$\omega$ to the external influences $\mu$ and the nonlinearity $\lambda$, providing insights into how external factors and the platform's feedback mechanisms influence the stability and characteristics of user behavior patterns.

Implementing this metaphorical model faces challenges, notably in translating the abstract mathematical concepts of spin solitons into tangible strategies for digital health platforms. Key among these challenges is the identification of relevant parameters that accurately reflect user behavior dynamics and the external influences within digital health ecosystems.

Future research could explore empirical validation of this model through data analysis and simulations, aiming to identify stable behavior patterns and their response to interventions. This approach could offer novel insights into designing digital health interventions that promote sustainable health improvement behaviors, ultimately contributing to better health outcomes and user engagement on digital platforms.

In conclusion, the metaphorical application of spin solitons to model user behavior in digital health platforms presents a promising avenue for understanding and fostering stable, beneficial health improvement behaviors. By exploring the nonlinear dynamics of user engagement and the influence of external factors, we can develop more effective strategies for health promotion and misinformation management in the digital health landscape.

Incorporating the concept of tellurium nanoparticles (TeNPs) into the exploration of spin soliton solutions within the context of nonlinear Schrödinger equations (NLS) to model the dynamics of fake news diffusion offers an intriguing multidisciplinary approach. The NLS equation, known for its soliton solutions that maintain their shape during propagation, provides a robust framework for understanding how information or misinformation waves interact in a digital environment.

## 2.11 Nonlinear Schrödinger Equation (NLS Equation) with Tellurium Nanoparticles

The generalized NLS equation can be extended to include the effects of TeNPs on the propagation of information waves by introducing an additional term that represents the interaction between the information field and the TeNPs:

$$i\frac{\partial \psi}{\partial t} + \frac{1}{2}\frac{\partial^2 \psi}{\partial x^2} + |\psi|^2 \psi + V(x)\psi = 0$$

Here, $V(x)$ represents the potential created by the presence of TeNPs, which can affect the propagation of the information wave $\psi(x, t)$.

Soliton Solution Ansatz and Analysis

To find soliton solutions in this modified NLS equation, we use the ansatz:

$$\psi(x, t) = A \operatorname{sech}(a(xvt)) e^{i(bx\omega t)}$$

Substituting this ansatz into the modified NLS equation and performing a series of algebraic manipulations yield conditions that the parameters $A$, $a$, $b$, $\omega$, and $v$ must satisfy. The presence of $V(x)$ introduces new conditions that reflect the influence of TeNPs on the propagation and stability of the information wave.

## 2.12 TeNPs Influence on Information Wave Propagation

The interaction term $V(x)\psi$ can be modeled based on the specific properties of TeNPs, such as their size, shape, and concentration, which determine how they interact with the information wave. This term could represent the adsorption or scattering of misinformation waves by TeNPs, effectively filtering or altering the misinformation as it propagates through the digital medium.

Modeling TeNPs Influence, Accurately modeling the influence of TeNPs on the propagation of information waves requires a deep understanding of both the physical properties of TeNPs and the characteristics of information waves in digital platforms.

Parameter Identification,Identifying the appropriate parameters for $V(x)$ that accurately represent the interaction between TeNPs and misinformation waves is crucial. This involves understanding the mechanisms through which TeNPs can adsorb or alter misinformation.

Practical Implementation,Translating this theoretical model into practical strategies for mitigating fake news on digital platforms poses significant challenges. It involves not only the development of algorithms that can mimic the behavior of TeNPs in adsorbing misinformation but also ethical considerations regarding information filtering and censorship.

Further research could explore more sophisticated models that incorporate the dynamic behavior of TeNPs and their interaction with various types of misinformation. Experimental studies could validate these models and assess their effectiveness in real-world scenarios. Additionally, exploring the ethical implications and developing transparent and accountable mechanisms for misinformation management will be critical for the practical application of these concepts in digital health platforms.

In summary, the metaphorical application of TeNPs and the exploration of spin soliton solutions in the context of the NLS equation offer a novel approach to understanding and managing the diffusion of fake news. This interdisciplinary framework combines concepts from materials science and nonlinear dynamics to propose innovative strategies for mitigating misinformation in the digital age, albeit with significant challenges and considerations for practical implementation.

To delve into the detailed mathematics and computational processes involved in exploring spin soliton solutions within the context of nonlinear partial differential equations for modeling the spread of fake news, let's consider a hypothetical scenario where tellurium nanoparticles (TeNPs) are used to adsorb misinformation in a digital health platform. This scenario involves applying the concept of spin solitons metaphorically to understand the strategic behaviors of first movers (originators of information) and second movers (spreaders or reactors to the information) within the nonlinear dynamics of information spread.

The dynamics of information spread, influenced by the actions of first and second movers, can be modeled using a modified version of the nonlinear Schrödinger equation (NLS), which is known to possess spin soliton solutions. The equation can be represented as:

$$i\frac{\partial \psi}{\partial t} + \Delta \psi + |\psi|^2 \psi = V(x,t)\psi$$

Here, $\psi(x,t)$ represents the complex field associated with the spread of information, $\Delta$ denotes the Laplacian operator accounting for the diffusion of information across the network, and $V(x,t)$ represents the potential induced by the presence of TeNPs, which can adsorb or alter misinformation.

## 2.13 Soliton Solution Ansatz for First and Second Movers

For the first mover (A), the soliton solution ansatz might take the form:

$$\psi_A(x,t) = A_A \operatorname{sech}(B_A(xC_At))e^{i(k_A x \omega_A t)}$$

For the second mover (B), dependent on the first mover's influence, the ansatz could be:

$$\psi_B(x,t) = A_B \operatorname{sech}(B_B(xC_Bt))e^{i(k_B x \omega_B t)}$$

## 2.14 Incorporating Tellurium Nanoparticles

The potential $V(x,t)$ introduced by TeNPs can be modeled as a function that selectively interacts with the misinformation components of the information field $\psi$. This interaction might depend on the characteristics of the misinformation and the properties of the TeNPs, such as size, shape, and surface chemistry.

1. Substituting the Ansatz: Insert the soliton solutions $\psi_A$ and $\psi_B$ into the modified NLS equation.

2. Separation of Variables: Perform a separation of variables to isolate terms involving spatial and temporal components.

3. Parameter Conditions: Equate coefficients of like terms to derive conditions on the parameters $A$, $B$, $C$, $k$, and $\omega$ for both first and second movers.

4. Influence of TeNPs: Analyze how the potential $V(x,t)$ alters the conditions for soliton solutions, reflecting the adsorption of misinformation by TeNPs.

Model Complexity: The interaction between the complex field $\psi$ and the potential $V(x,t)$ introduced by TeNPs adds significant complexity, necessitating sophisticated numerical methods for solution and analysis. Parameter Estimation: Determining the appropriate parameters for $V(x,t)$ that accurately model the TeNPs' effect on misinformation requires empirical data and advanced fitting techniques. Interpretation and Validation: The metaphorical application of spin solitons and TeNPs in this context requires careful interpretation, and the model's predictions should be validated against real-world data on information spread and the efficacy of misinformation countermeasures.

This mathematical exploration provides a theoretical foundation for understanding how TeNPs might be used to adsorb fake news within a digital health platform, using the metaphor of spin solitons. The model highlights the complex interplay between information spread dynamics, the strategic behaviors of information movers, and the potential role of nanotechnology in mitigating misinformation. However, the practical application of these concepts would require further interdisciplinary research and empirical validation.

The exploration of spin soliton solutions in the context of serious games, particularly considering the strategic behaviors of first movers and second movers, requires a sophisticated theoretical approach that integrates elements of game theory with nonlinear partial differential equations. Here, we delve into a conceptual framework that applies the idea of spin solitons metaphorically to model the dynamics of information spread, such as fake news, within a digital platform, utilizing the unique properties of tellurium nanoparticles (TeNPs) for the adsorption scenario.

1. State Variable: Consider a state variable $S(x,t)$, where $x$ represents the strategic space, and $t$ denotes time or the progression of the game rounds.

2. Nash Equilibrium: The strategic equilibrium, where both first movers (A) and second movers (B) adopt their best response strategies, can be analyzed through the lens of Nash Equilibrium.

3. Hooke's Law Analogy: The concept of Hooke's law is metaphorically applied to represent the 'elasticity' or adaptability of strategies within the game, indicating how strategies may 'stretch' or 'compress' in response to external pressures or opportunities.

## 2.15 Nonlinear Partial Differential Equation

The dynamics of the game can be expressed through a nonlinear partial differential equation:

$$\frac{\partial S}{\partial t} + \alpha S \frac{\partial S}{\partial x} = \beta \frac{\partial^2 S}{\partial x^2} + G(S,x,t)$$

Here, $G(S, x, t)$ is a function representing specific rules or conditions of the game, incorporating the advantages of first movers and the strategic responses of second movers.

**2.16 Soliton Solution Ansatz**

A soliton solution ansatz is proposed to find stable patterns of strategic behavior:

$$S(x, t) = A \operatorname{sech}^2(B(xCt))$$

where $A$ denotes the amplitude of the soliton, representing the intensity of strategic actions; $B$ indicates the width of the soliton, reflecting the range of strategies employed; and $C$ signifies the soliton's velocity, capturing the speed at which strategies evolve over time.

**2.17 Strategies of First and Second Movers**

First Mover, The first mover's strategy $S_F(x, t)$ is expressed through the soliton solution, with the first mover's advantage represented by $G_F(S, x, t)$, integrated into the nonlinear partial differential equation.

Second Mover, The second mover's strategy $S_S(x, t)$, also formulated as a soliton solution, is adjusted based on the first mover's actions, represented by $G_S(S, x, t, S_F)$, and incorporated into the equation.

1. Differentiating the Ansatz: Calculate the time and spatial derivatives of $S_F(x, t)$ and $S_S(x, t)$.

2. Substituting into the Equation: Insert the differentiated results into the nonlinear partial differential equation to derive conditions for the strategies of first and second movers.

3. Analyzing Soliton Solution Properties: Analyze the derived conditions to determine the parameters $A$, $B$, and $C$ for the soliton solutions of first and second movers, thereby understanding the characteristics of each strategy.

This approach, while offering a rich conceptual framework for analyzing the strategic dynamics of serious games through the metaphorical application of spin solitons, requires careful consideration for practical application. The specific forms of $G_F$ and $G_S$ need to be meticulously defined, and the physical interpretation of soliton solutions should be thoughtfully considered. Validation of the model's predictions against empirical data from game play and player feedback is essential to ensure the model's relevance and applicability to real-world scenarios.

# 3. Fundamentals of Plasmon Resonance

To understand the theoretical background and calculation process related to tellurium nanoparticles and plasmonic phenomena, we start from the basic principles of physics. Plasmonics is a phenomenon that arises from the interaction between light (electromagnetic waves) and free electrons in metals. Localized surface plasmon resonance (LSPR) in metal nanoparticles is particularly intriguing and depends on the size, shape, and dielectric properties of the surrounding medium.

The condition for plasmon resonance is defined by the resonant interaction between electromagnetic waves and free electrons on the surface of nanoparticles. This resonance condition is often explained using the Drude model, where the dielectric function $\epsilon(\omega)$ of the metal is given by:

$$\epsilon(\omega) = \epsilon_\infty \frac{\omega_p^2}{\omega^2 + i\omega\gamma}$$

where, $\epsilon_\infty$ is the high-frequency limit of the dielectric constant, $\omega_p$ is the plasma frequency, $\omega$ is the angular frequency of the incident light, $\gamma$ is the damping constant.

# 4. Localized Surface Plasmon Resonance in Tellurium Nanoparticles

Localized surface plasmon resonance in tellurium nanoparticles is observed through strong absorption and scattering of electromagnetic waves at specific resonance frequencies. This resonance frequency varies depending on the shape of the nanoparticles and the dielectric properties of the surrounding medium. The resonance frequency is determined by the relationship between the dielectric function of the nanoparticles and that of the surrounding medium. Mie theory provides analytical solutions for scattering and absorption by spherical particles and is expressed as follows:

$$Q_{\text{abs}} = \frac{4\pi k}{\lambda} \operatorname{Im} \left\{ \frac{\epsilon_{\text{particle}} \epsilon_{\text{medium}}}{\epsilon_{\text{particle}} + 2\epsilon_{\text{medium}}} \right\}$$

$$Q_{\text{sca}} = \frac{8\pi k^4}{3\lambda^4} \left| \frac{\epsilon_{\text{particle}} \epsilon_{\text{medium}}}{\epsilon_{\text{particle}} + 2\epsilon_{\text{medium}}} \right|^2$$

where, $Q_{\text{abs}}$ is the absorption efficiency, $Q_{\text{sca}}$ is the scattering efficiency, $k$ is the radius of the sphere, $\lambda$ is the wavelength of the incident light, $\epsilon_{\text{particle}}$ and $\epsilon_{\text{medium}}$ are the dielectric functions of the nanoparticles and the surrounding medium, respectively.

Before delving into specific calculations, experimental data regarding the specific physical and chemical properties of tellurium nanoparticles (e.g., plasma frequency $\omega_p$ and damping constant $\gamma$) are required. Using these parameters, the resonance frequency of localized surface plasmons under specific conditions (e.g., particle size, shape, dielectric properties of the surrounding medium) can be calculated based on the above equations.

# 5. To the Drude Model

To understand and control the diffusion and impact of fake news, it is an effective approach to build a theoretical framework using the Drude model and plasmonics phenomena. In this section, we will delve into understanding the localized surface plasmon and surface plasmon using the Drude model and explain in detail how it can be applied to the "adsorption" of fake news and strategies for preventing its diffusion.

The Drude model is a classical theory for analyzing the optical properties of metals, describing how free electrons in metals respond to external electromagnetic fields (light). This model is represented by the following equation:

$$\epsilon(\omega) = 1 \frac{\omega_p^2}{\omega^2 + i\omega\gamma}$$

where, $\epsilon(\omega)$ is the complex dielectric constant of the metal, $\omega_p$ is the plasma frequency of the metal (the natural frequency of collective oscillations of free electrons), $\omega$ is the angular frequency of the external electromagnetic wave, $\gamma$ is the damping constant (representing energy loss due to electron collisions).

## 6. Understanding Localized Surface Plasmon Resonance and Surface Plasmon Resonance

Localized surface plasmon resonance (LSPR) and surface plasmon resonance (SPR) in metal nanoparticles and nanostructures can be explained based on the Drude model. These resonance phenomena occur when the frequency of the external electromagnetic wave is close to the plasma frequency, efficiently exchanging energy between the electromagnetic wave and the free electrons in the metal. This energy exchange excites collective electron oscillations and leads to enhancement of localized electromagnetic fields near metal nanoparticles and nanostructures.

The diffusion of fake news and its resonance within social groups can be likened to localized surface plasmon resonance and surface plasmon resonance. In this analogy, fake news corresponds to the external electromagnetic wave, while individuals or subgroups within the social group correspond to metal nanoparticles or nanostructures. When fake news resonates with specific groups or individuals, the information is strongly accepted and rapidly disseminated.

## 7. Strategies for Converging Fake News Diffusion

By applying the Drude model and plasmonics phenomena to the management of fake news, strategies can be devised to converge the resonance and diffusion of information. Specifically, the following approaches can be considered:
1. Shift of Resonance Frequency: By changing conditions that make fake news more resonant (e.g., specific emotions or beliefs), the likelihood of information resonance can be reduced. 2. Increase in Damping Constant: Through education and awareness campaigns targeted at individuals or groups susceptible to the influence of fake news, resistance to information acceptance and diffusion can be increased. 3. Management of External Electromagnetic Waves (Information): By providing accurate information and promptly correcting fake news, the influx of information from external sources can be managed, preventing the spread of misinformation.

## 8. Drude Model

To explain the analogy between fake news diffusion and plasmonics phenomena, especially when associating with the Drude model, it's necessary to understand how the Drude model describes the optical properties of metals. Then, we'll explore how this physical model can be applied to the metaphor of fake news diffusion.

The Drude model describes how free electrons in metals respond to external electromagnetic waves (light) from a classical perspective. The complex dielectric constant $\epsilon(\omega)$ of a metal is given by the following equation:

$$\epsilon(\omega) = \epsilon_\infty \frac{\omega_p^2}{\omega^2 + i\omega\gamma}$$

where, $\epsilon_\infty$ is the dielectric constant at infinite frequency, $\omega_p$ is the plasma frequency (the natural frequency of collective electron oscillations), $\omega$ is the angular frequency of the external electromagnetic wave, $\gamma$ is the damping constant (representing the rate of energy loss due to collisions).

## 9. Localized Surface Plasmon Resonance (LSPR)

Localized surface plasmon resonance (LSPR) in metal nanoparticles occurs when the frequency of the external electromagnetic wave is close to the plasma frequency. At this point, there is an enhancement of the electromagnetic field near the metal nanoparticle. The resonance condition occurs when the dielectric constant of the nanoparticle satisfies a specific relationship with the dielectric constant of the surrounding medium. For spherical nanoparticles, the resonance condition is simplified as:

$$\text{Re}\left\{\epsilon(\omega)\right\} = -2\epsilon_\text{m}$$

where $\epsilon_\text{m}$ is the dielectric constant of the surrounding medium.

### 9.1 Application to the Metaphor of Fake News Diffusion

To apply this physical model to the context of fake news diffusion, fake news is likened to external electromagnetic waves, while individuals or subgroups within the social group are likened to metal nanoparticles. The resonance condition corresponds to fake news "resonating" with specific groups or individuals, amplifying its impact. In this analogy, the following computational processes can be assumed:

1. Identification of Resonance Conditions: Identify the characteristics of social groups or individuals most susceptible to resonance with fake news. This corresponds to identifying the "characteristic frequency" of fake news, analogous to the plasma frequency.

2. Amplification of Impact: Model how fake news is accepted and propagated within specific groups based on resonance conditions. At this stage, the amplification of the electromagnetic field due to resonance corresponds to an increase in the receptiveness and diffusion rate of fake news.

3. Derivation of Convergence Strategies: Calculate strategies to suppress resonance with fake news. This may involve altering resonance conditions (e.g., changing individual perceptions through education) or counteracting the effects of resonance (e.g., neutralizing the impact of fake news with accurate information).

## 10. Understanding "Resonance" of Fake News

When devising convergence strategies for the diffusion of fake news, directly applying mathematical equations or computational processes is challenging due to the metaphorical nature of the approach. When linking phenomena in plasmonics with issues in social sciences, conceptual frameworks and analogies are primarily used rather than mathematical models. However, it's possible to quantitatively understand some fundamental concepts when considering theoretical approaches to address the diffusion of fake news.

By likening the resonance of fake news with specific social groups or individuals to the resonance phenomena in plasmonics, the following steps can be considered:

1. Identification of Resonance Conditions: In physics, resonance conditions are determined by the plasma frequency defined in models like the Drude model. In the context of fake news, "resonance conditions" refer to the characteristics or belief systems of social groups where fake news exerts the most influence.

2. Quantification of Impact: In plasmonics, the amplification of the electromagnetic field due to resonance can be expressed mathematically. In the case of fake news, the "strength" of resonance may be quantified by changes in the diffusion rate or receptiveness to fake news. For example, measuring increases in shares or comments on social media can estimate the strength of resonance.

## 11. Development of Convergence Strategies

When devising convergence strategies to counteract the diffusion of fake news, the following elements should be considered:

1. Modification of Resonance Conditions: In physics, resonance frequencies can be altered by external perturbations. In the context of fake news, changing the "resonance conditions" can be achieved by enhancing people's receptiveness to information and critical thinking abilities. This may involve educational programs or awareness campaigns.

2. Introduction of Damping Mechanisms: In plasmonics, resonance damping occurs due to material losses or scattering. For combating fake news, introducing "damping" mechanisms can be done by spreading the results of fact-checking or removing misinformation on social media platforms.

## 12. Equation of the Drude Model

In plasmon resonance based on the Drude model, the resonance frequency depends on the interaction between the oscillation of free electrons within the metal and external electromagnetic waves. To understand the shift in resonance frequency, one must consider the basic equation of the Drude model and the conditions for plasmon resonance.

The complex dielectric function of a metal according to the Drude model is given by the following equation:

$$\epsilon(\omega) = \epsilon_\infty \frac{\omega_p^2}{\omega^2 + i\omega\gamma}$$

Here, $\epsilon(\omega)$ is the complex dielectric function of the metal at frequency $\omega$, $\epsilon_\infty$ is the dielectric constant at high frequency limit, $\omega_p$ is the plasma frequency, $\omega$ is the angular frequency of the external electromagnetic wave, $\gamma$ is the damping constant of the free electrons.

## 13. Plasmon Resonance Frequency

For metal nanoparticles, the condition for local plasmon resonance (LSPR) occurs when the dielectric function of the nanoparticle satisfies a specific relationship with the dielectric constant of the surrounding medium. The simplified resonance condition for spherical nanoparticles is expressed as follows:

$$\text{Re}\left\{\epsilon(\omega)\right\} = -2\epsilon_m$$

Here, $\epsilon_m$ is the dielectric constant of the surrounding medium.

## 14. Shift in Resonance Frequency

The shift in resonance frequency occurs due to changes in the plasma frequency $\omega_p$. The plasma frequency is given by the following equation:

$$\omega_p = \sqrt{\frac{Ne^2}{m\epsilon_0}}$$

Here, $N$ is the number of free electrons per unit volume, $e$ is the charge of an electron, $m$ is the mass of an electron, $\epsilon_0$ is the vacuum permittivity.

To calculate the shift in resonance frequency, changes in $N$, $m$, or $\epsilon_0$ must be considered. In practice, changes in $N$ are common due to variations in nanoparticle size, shape, surrounding medium, or doping.

For example, if the density of free electrons $N$ increases due to nanoparticle doping, the plasma frequency $\omega_p$ increases, resulting in a higher resonance frequency. This shift can be calculated by substituting the new value of $N$ into the above plasma frequency equation.

Thus, the shift in resonance frequency is calculated based on changes in the physical and chemical properties of nanoparticles. By controlling this shift, the optical properties of nanoparticles can be adjusted. In the analogy of fake news, this "shift" suggests changes in how information is received and reacted to, based on changes in people's perceptions, belief systems, or the context of information.

## 15. Complex Dielectric Function in the Drude Model

In the Drude model, the damping constant $\gamma$ represents the energy loss of free electrons within the metal when they react to external electromagnetic waves. This constant is closely related to the optical response of the metal and plays an important role in the complex dielectric function of the metal.

The complex dielectric function of a metal based on the Drude model is given by the following equation:

$$\epsilon(\omega) = \epsilon_\infty \frac{\omega_p^2}{\omega^2 + i\omega\gamma}$$

Here, $\epsilon(\omega)$ is the complex dielectric function of the metal at frequency $\omega$. $\epsilon_\infty$ is the dielectric constant at infinite frequency. $\omega_p$ is the plasma frequency. $\omega$ is the angular frequency of the external electromagnetic wave. $\gamma$ is the damping constant.

## 16. Effects of Increasing Damping Constant $\gamma$

The damping constant $\gamma$ represents the rate of energy loss of free electrons within the metal when they react to external electromagnetic waves. A larger value of this constant leads to greater energy dissipation within the metal, resulting in a decrease in the sharpness of resonance. Specifically, increasing the damping constant has the following effects:

1. Broadening of Resonance: A larger $\gamma$ leads to a broader resonance peak. This indicates higher energy loss and results in a more "blurred" resonance state.

2. Decrease in Resonance Intensity: The height (intensity) of the resonance peak decreases as $\gamma$ increases. This indicates weaker enhancement of the electromagnetic field due to faster energy dissipation.

## 17. Derivation of Equations Regarding Increase in Damping Constant

To demonstrate the specific effects of increasing the damping constant $\gamma$ on resonance characteristics, consider the behavior of the complex dielectric function near the resonance frequency. Near the resonance frequency, $\omega \approx \omega_p$, so the dielectric function can be approximated as follows:

$$\epsilon(\omega) \approx \epsilon_\infty \frac{\omega_p^2}{\omega_p^2 + i\omega_p\gamma}$$

Both the real and imaginary parts of the complex dielectric function at the resonance frequency depend on the damping constant $\gamma$. In particular, the imaginary part $\text{Im}(\epsilon)$ is directly related to the sharpness of resonance and can be expressed as follows:

$$\text{Im}(\epsilon) = \frac{\omega_p^3 \gamma}{(\omega_p^2)^2 + (\omega_p\gamma)^2}$$

From this equation, it is evident that increasing $\gamma$ leads to changes in the value of $\text{Im}(\epsilon)$ and alters the resonance characteristics. Specifically, increasing $\gamma$ reduces the sharpness of resonance, resulting in a broader resonance peak.

Through such analysis, it is possible to adjust the optical properties of metal nanoparticles. In the context of fake news, the concept of "damping" can be metaphorically applied to strategies (such as promoting critical thinking or widespread fact-checking) aimed at reducing the impact of fake news.

## 18. Influence of External Electromagnetic Waves in the Drude Model

When discussing the control of external electromagnetic waves in the context of the Drude model, in physics, it refers to controlling the behavior of electromagnetic waves (light) incident on a metal from the outside. The complex dielectric function of the metal determines how the metal absorbs, reflects, and transmits these electromagnetic waves. In the context of the spread of fake news, managing these "external electromagnetic waves" metaphorically represents managing the influx of information or news.

The complex dielectric function of a metal according to the Drude model is given by:

$$\epsilon(\omega) = \epsilon_\infty - \frac{\omega_p^2}{\omega^2 + i\omega\gamma}$$

where: $\epsilon(\omega)$ is the complex dielectric function of the metal at frequency $\omega$, $\epsilon_\infty$ is the dielectric constant at high frequencies, $\omega_p$ is the plasma frequency, $\omega$ is the angular frequency of the external electromagnetic wave, $\gamma$ is the damping constant.

Managing external electromagnetic waves (information) in this context means adjusting the frequency $\omega$ of incident waves to control the optical response of the metal. Specifically, it is possible to vary the metal's reflectance, transmittance, and absorbance based on frequency.

The reflectance $R$ of a metal can be calculated using the complex dielectric function as follows:

$$R = \left| \frac{\sqrt{\epsilon(\omega)} - 1}{\sqrt{\epsilon(\omega)} + 1} \right|^2$$

By substituting the Drude model's dielectric function into this equation, the reflectance $R$ for a specific $\omega$ (frequency of the external electromagnetic wave) can be computed.

# 19. Managing "External Electromagnetic Waves" in the Context of Fake News

Managing "external electromagnetic waves (information)" in the context of the spread of fake news involves controlling the quality and circulation of information. This includes strategies such as:

Monitoring Information Sources: Prioritizing information from reliable sources and filtering out inaccurate sources. Fact-Checking and Verification: Verifying published information and debunking misinformation. Education and Awareness: Enhancing the ability of the general public to critically evaluate information and identify fake news.

By implementing these strategies, the spread of fake news can be curtailed, and the quality of information in society can be improved. By using the principles of managing external electromagnetic waves in the physics Drude model as an analogy for managing the circulation of fake news and information, insights can be gained to more effectively control the circulation of information.

# 20. Modeling Adsorption in Nanoparticles

When using tellurium nanoparticles as a metaphor for the "adsorption" of fake news, applying direct mathematical models is challenging due to the abstract nature of the concept. However, to concretize this metaphor, it is possible to refer to chemical and physical approaches that model the process of nanoparticles adsorbing specific molecules or substances.

Common models describing adsorption of substances include the Langmuir adsorption isotherm and the Freundlich adsorption isotherm. These equations capture different aspects of the adsorption process, assuming either a monolayer formed by adsorbate on a fixed surface (Langmuir) or allowing for multilayer adsorption (Freundlich).

## 20.1 Langmuir Adsorption Isotherm

The Langmuir model assumes constant adsorption sites and monolayer adsorption, and it is represented by the following equation:

$$\frac{\theta}{P} = \frac{1}{K_L} + \frac{\theta}{K_L C_{\max}}$$

where: $\theta$ is the surface coverage (the proportion of the surface covered by the adsorbate), $P$ is the pressure or concentration of the adsorbate, $K_L$ is the Langmuir constant (indicating the affinity of adsorption), $C_{\max}$ is the maximum adsorption capacity.

## 20.2 Application to "Adsorption" of Fake News

When applying the Langmuir model to the metaphor of "adsorption" of fake news, it can be interpreted as follows:

$\theta$: The proportion of a specific group or individual influenced by fake news. $P$: The "concentration" of fake news or the degree of exposure through the media. $K_L$: The receptivity to fake news or the "affinity" of specific groups to be attracted to fake news. $C_{\max}$: The "maximum capacity" of fake news acceptable to specific groups.

This metaphor provides a theoretical framework to understand the dynamics of the spread and acceptance of fake news, indicating to what extent specific groups are "adsorbed" to fake news. However, this approach is metaphorical and cannot directly apply actual numerical values or computational results. Quantitatively analyzing the impact of fake news requires methods and data analysis from social sciences.

# 21. Snell's Law

When applying Snell's law as a metaphor for the "refraction" of information and resonance, the expansion of the concept is based on the equations of light refraction in physics, but there are no specific formulas or calculations directly applicable to fake news. However, to understand this analogy, let's first review the basics of Snell's law and then consider how the concepts of "refraction" and "resonance" in information propagation can be interpreted.

Snell's law describes the relationship between the angle of incidence and the angle of refraction when light refracts at the boundary between different media. The equation is represented as follows:

$$n_1 \sin(\theta_1) = n_2 \sin(\theta_2)$$

where: $n_1$ is the refractive index of the first medium, $\theta_1$ is the angle of incidence (the angle at which light enters the boundary from the first medium), $n_2$ is the refractive index of the second medium, $\theta_2$ is the angle of refraction (the angle at which light progresses within the second medium).

The "refraction" of information as it propagates between different cultures or social groups refers to the phenomenon where information is interpreted differently based on the assumptions and values held by the recipient. In this metaphor, the "refractive index" represents the flexibility and receptivity of social groups to interpret information, showing how information transforms differently among different groups.

The "resonance" of information refers to the phenomenon where specific information strongly resonates with existing beliefs or emotions of the recipient, eliciting strong reactions. In this metaphor, the resonance phenomenon of light in tellurium nanoparticles is likened to situations where information holds particular influence within specific groups.

To apply this metaphor to concrete analysis and strategies, the following steps can be considered:

1. Evaluation of "Refractive Index": Evaluate how different social groups are likely to interpret information and understand their "refractive index". 2. Adaptation of Information: Adjust the presentation of information to align with the "refractive index" of the target group when communicating information to different groups. 3. Promotion of Resonance: Devise ways to ensure that information resonates strongly within specific groups by carefully crafting its content and delivery.

Incorporating tellurium nanoparticles and plasmonic phenomena into the context of incomplete information games and the repeated dilemma means constructing a conceptual framework to understand the diffusion of fake news and its countermeasures, rather than providing direct formulas or calculation processes. However, it's possible to formalize this concept using the basic models of game theory. Here, we illustrate one way to model the interaction between players in an incomplete information game.

## 22. Incorporating Tellurium Nanoparticles and Plasmonic Phenomena

In incomplete information games, players do not have complete information about the types or choices of other players. In the context of the repeated dilemma, players repeatedly make the same choices and adjust their strategies based on past choices. In this context, players can be, for example, disseminators and consumers (or debunkers) of fake news.

Player 1 (Disseminator) Strategy: Spread (S) fake news or not (NS). Player 2 (Consumer/Debunker) Strategy: Verify (V) information or ignore (I) it.

## 23. Payoff Functions

Player payoffs depend on the chosen strategies and the opponent's strategies. Below is a simple table representing the payoffs.

| Player\Strategy | $V$ | $I$ |
|---|---|---|
| $S$ | $(-a, b)$ | $(c, -d)$ |
| $NS$ | $(0, 0)$ | $(0, 0)$ |

where: $a$ is the disseminator's loss when fake news is verified. $b$ is the gain for consumers/debunkers when fake news is verified. $c$ is the disseminator's gain when fake news is spread and ignored. $d$ is the loss for consumers/debunkers when fake news is ignored.

To incorporate tellurium nanoparticles and plasmonic phenomena into this model, we need to consider the "resonance" and "refraction" of information. This functions as a metaphor to show how fake news is accepted and spread within specific groups.

Resonance: If fake news is strongly accepted within specific groups, the gain $c$ (for disseminators) will increase. Refraction: When information propagates between different groups, the gains $b$ and losses $d$ (for consumers/debunkers) will vary.

One way to analyze this game is to find Nash equilibria. In a Nash equilibrium, no player can increase their payoff by changing their strategy. In this model, analyzing how equilibria change with different parameter values $a, b, c, d$ can provide insights into the dynamics.

This framework serves as a starting point to understand the dynamics of fake news diffusion and countermeasures, but capturing real social phenomena requires more complex models and additional elements.

## 24. Incorporating Tellurium Nanoparticles and Plasmonic Phenomena

In the context of the repeated dilemma in incomplete information games, incorporating the concepts of the Drude model and tellurium nanoparticles into the context of fake news absorption and arbitration does not directly apply physical phenomena to game theory with specific formulas or calculation processes. However, it's possible to use these concepts metaphorically to model the mechanism of fake news diffusion. Below, we present one way to formalize this idea.

**Players**: Player A (Disseminator): Chooses to spread (D) or not spread (ND) fake news. Player B (Receiver): Chooses to accept (A) or verify (V) the news.

**Player A's Strategy**: D or ND **Player B's Strategy**: A or V **Payoff Functions**: Player A's payoff: $u_A(D, A) = a, u_A(D, V) = -b, u_A(ND, \_) = 0$ Player B's payoff: $u_B(D, A) = -c, u_B(D, V) = d, u_B(ND, \_) = 0$ where $a$, $b$, $c$, and $d$ are positive constants representing the gains or losses for players' strategies.

To model the "absorption" effect of fake news by tellurium nanoparticles, we introduce additional parameters that influence player B's strategy choice. This is termed as the "absorption rate" $\alpha$. The probability of player B choosing A increases due to the persuasiveness or resonance of fake news, represented by $\alpha$.

Similarly, to model the "arbitration" effect, we introduce an additional payoff $\beta$ when player B chooses V. This represents the influence of external verification mechanisms or fact-checking.

## 25. Adjusted Payoff Functions

Player A's payoff: $u_A(D, A) = a\alpha, u_A(D, V) = -b, u_A(ND, \_) = 0$ Player B's payoff: $u_B(D, A) = -c, u_B(D, V) = d + \beta, u_B(ND, \_) = 0$

## 26. Deriving Nash Equilibrium

To find Nash equilibria, we search for strategy combinations that maximize the payoff for each player. In this model, the optimal strategies of players vary depending on the values of $\alpha$ and $\beta$.

This modeling allows us to understand how fake news "absorption" and arbitration mechanisms influence player strategy choices. However, this model is merely a conceptualization using metaphors and does not fully capture the complexity of fake news diffusion and people's reactions. For a more precise analysis, real data and detailed psychological and sociological studies are required.

In the context of the repeated dilemma in incomplete information games dealing with the diffusion of fake news, incorporating the concepts of Snell's law and tellurium nanoparticles into the scenario leads to applying the logic of fake news "absorption" and game theory "arbitration" metaphorically, and even introducing specific patterns of societal group reactions and the notion of local plasmons, thus applying physical phenomena metaphorically to game theory. While it's not directly possible to convert physical processes into formulas of game theory, it's possible to construct models to understand the dynamics of fake news diffusion and its countermeasures using these concepts.

**Players**: Fake news disseminator (A) and societal group (B). **Strategies**: Strategy of A: Spread fake news (D) or not (ND). Strategy of B: Accept (A) or verify (V) the fake news.

## 27. Payoff Functions

**Player A's Payoff**: $u_A(D, A) = a, u_A(D, V) = -b, u_A(ND, \_) = 0$ **Player B's Payoff**: $u_B(D, A) = -c, u_B(D, V) = d, u_B(ND, \_) = 0$

Using Snell's law as a metaphor for the refraction of information and utilizing the local plasmon resonance of tellurium nanoparticles as a metaphor for the "absorption" of fake news. Factors influencing the diffusion and acceptance of fake news can be modeled as parameters corresponding to the intensity and frequency of local plasmon resonance.

**Fake News "Absorption"**: The degree to which fake news is accepted by the societal group is represented by a parameter $\alpha$ corresponding to the intensity of local plasmon resonance. **Societal Group Reaction Patterns**: The effect of societal group's verification strategy on fake news is represented by a parameter $\beta$ corresponding to the frequency shift of local plasmon resonance.

Modify the payoff functions of players A and B using $\alpha$ and $\beta$ and then find the Nash equilibrium.

**Modified Payoff Function for Player A**: $u_A(D, A) = a\alpha, u_A(D, V) = -b\beta, u_A(ND, \_) = 0$ **Modified Payoff Function for Player B**: $u_B(D, A) = -c\alpha, u_B(D, V) = d\beta, u_B(ND, \_) = 0$

To find the Nash equilibrium, we look for strategies where each player maximizes their payoff against the other's strategy. In this process, the optimal strategies change depending on the values of $\alpha$ and $\beta$.

When considering the role of the ultimate receiver of information as the "mover" in the incomplete information game of fake news diffusion, metaphorically applying the concepts of Surface Plasmon Resonance (SPR) and tellurium nanoparticles provides insights into understanding the mechanisms of acceptance and reaction to fake news. This approach offers a framework for examining how information influences receivers and spreads within societal groups.

## 28. Metaphor of Surface Plasmon Resonance (SPR)

Surface Plasmon Resonance is a phenomenon where the collective oscillation of free electrons inside a metal resonates with light waves on the metal surface. This resonance occurs strongly under specific conditions (e.g., angle or frequency of incident light), causing localized amplification of the electromagnetic field near the metal surface.

In the context of fake news, SPR can serve as a metaphor for how information receivers interact with fake news. Under specific conditions, fake news can "resonate" with receivers, amplifying its impact locally. This "resonance" may occur particularly strongly when the receiver's existing beliefs, emotions, or values align with the content of fake news.

## 29. Metaphor of Tellurium Nanoparticles

Tellurium nanoparticles are substances capable of effectively inducing localized plasmon resonance due to their characteristics. The effect of localized plasmon resonance in tellurium nanoparticles can be metaphorically applied to understand how fake news spreads and amplifies within societal groups.

In this metaphor, tellurium nanoparticles represent socially sensitive groups or subgroups particularly susceptible to fake news, depicting how these groups interact with fake news and amplify its influence. When fake news resonates with the values or beliefs of specific groups, the information spreads rapidly within the group, potentially exerting significant influence on the group's opinions and actions.

## 30. Modeling the Second Mover:Tellurium Nanoparticles and LSPR Basics

When modeling the role of the second mover, or the re-disseminator of information, in the incomplete information game of fake news diffusion using tellurium nanoparticles, rather than applying direct physical equations, the phenomenon of Localized Surface Plasmon Resonance (LSPR) of tellurium nanoparticles is used as a metaphor for the re-dissemination of fake news. Here, we explain the concept of modeling based on this metaphorical approach and propose a formula to represent the "efficiency" of re-dissemination.

Localized Surface Plasmon Resonance in tellurium nanoparticles is a phenomenon where the resonance of surface electrons of nano-sized particles occurs when light of a specific frequency is incident on them, causing localized amplification of the electromagnetic field. This resonance leads to a significant enhancement of the electromagnetic field near the nanoparticles.

When modeling the behavior of the second mover, the effect of LSPR is likened to the "efficiency" of the re-dissemination of fake news. The efficiency with which the second mover re-diffuses fake news depends on the extent to which the content of the fake news resonates with the beliefs or values of the second mover.

The re-dissemination efficiency $E$ of fake news by the second mover can be defined as follows:

$$E = \frac{1}{1 + e^{-k(R \cdot CS)}}$$

Here, $R$ represents the coefficient for the persuasiveness or influence of the fake news. $C$ represents the resonance coefficient for the content of the fake news to the second mover, i.e., how much the existing beliefs or values of the second mover align with the content of the fake news. $S$ represents the coefficient for the skepticism or critical thinking ability of the second mover. $k$ is a constant that adjusts the sensitivity of resonance. $e$ is the base of the natural logarithm.

In this model, as $R$ and $C$ increase, indicating stronger persuasiveness of the fake news and greater resonance with the beliefs or values of the second mover, the re-dissemination efficiency $E$ increases. Conversely, as $S$ increases, indicating greater skepticism or higher critical thinking ability of the second mover, the re-dissemination efficiency decreases.

This model provides a metaphorical framework for understanding the re-dissemination process of fake news. It demonstrates the significant role played by the second mover in the dissemination of fake news and provides insights for devising effective measures against fake news. Importantly, this model serves as a conceptual tool for analyzing the impact of fake news rather than providing direct computational processes.

## 31. Modeling the Third Mover, Tellurium Nanoparticles and LSPR Basics

When modeling the role of the third mover, i.e., the ultimate recipient of information, in the incomplete information game of fake news diffusion using tellurium nanoparticles, a metaphorical approach is taken instead of applying direct physical equations. Here, we construct a model using the concept of Localized Surface Plasmon Resonance (LSPR) in tellurium nanoparticles to illustrate how fake news is perceived and reacted to by the third mover.

Localized Surface Plasmon Resonance in tellurium nanoparticles is a phenomenon where the resonance of surface electrons of nano-sized particles occurs when light of a specific frequency is incident on them, causing localized amplification of the electromagnetic field. This resonance leads to a significant enhancement of the electromagnetic field near the nanoparticles.

When modeling the third mover as tellurium nanoparticles, the following elements are considered:

1. "Incident Light" of Fake News: Fake news can be seen as the light incident on the tellurium nanoparticles. The properties of this "light" (intensity, wavelength, polarization, etc.) correspond to the content, presentation, credibility, etc., of the fake news.

2. Resonance Condition: The resonance condition in tellurium nanoparticles corresponds to how the third mover reacts to fake news, i.e., how much they accept, reject, or verify the information. This resonance condition varies depending on the third mover's beliefs, values, biases, information verification abilities, etc.

3. Amplification of Electromagnetic Field: The amplification of the electromagnetic field resulting from resonance indicates the strength of the influence of fake news on the third mover. The stronger the resonance, the greater the influence of fake news.

Instead of applying direct physical equations, a metaphorical model framework is proposed. For example, the acceptance $A$ of fake news can be defined as follows:

$$A = \frac{R \cdot C}{1 + D}$$

Here, $R$ represents the coefficient for the credibility or persuasiveness of fake news. $C$ represents the resonance condition of the third mover, i.e., the extent to which the content of fake news aligns with the beliefs or values of the third mover. $D$ represents the coefficient for the difficulty of verification or lack of verification ability of the third mover.

This model provides a metaphorical framework for understanding how fake news is accepted by the third mover and the strength of its influence. Importantly, this model serves as a conceptual tool for analyzing the impact of fake news rather than providing direct computational processes.

## 32. Metaphor for Optical Properties of Tellurium Nanoparticles

When modeling the "transmittance" of the third mover in the incomplete information game of fake news diffusion using tellurium nanoparticles, a metaphorical approach is taken instead of applying direct physical equations. Here, we construct an abstract model using metaphors based on the optical properties of tellurium nanoparticles to illustrate how much fake news is accepted by the third mover.

Localized Surface Plasmon Resonance (LSPR) in tellurium nanoparticles exhibits strong absorption and scattering for light of specific frequencies, but may have high transmittance for light outside the resonance frequency. This optical property is used as a metaphor to represent how much fake news "passes through" to the third mover, i.e., is accepted.

The transmittance $T$ of the third mover is defined as the probability of fake news being accepted by them. This transmittance depends on both the characteristics of fake news and the beliefs and verification abilities of the third mover. It can be modeled in the following form:

$$T = \frac{1}{1 + e^{-(aI + bBcV)}}$$

Here, $I$ represents the parameters for the influence or persuasiveness of fake news. $B$ represents the strength of the third mover's existing beliefs or biases towards fake news. $V$ represents the third mover's information verification ability or willingness to verify. $a$, $b$, $c$ are coefficients adjusting the influence of $I$, $B$, $V$ respectively. $e$ is the base of the natural logarithm.

In this model, the larger the values of $I$ and $B$, i.e., the stronger the influence of fake news and the stronger the existing beliefs of the third mover, the higher the transmittance $T$. Conversely, the larger the value of $V$, i.e., the higher the verification ability of the third mover, the lower the transmittance.

This model provides a metaphorical framework for understanding how much fake news is accepted by the third mover, i.e., "passes through." By using metaphors based on the optical properties of tellurium nanoparticles instead of applying direct physical equations, insights into the dynamics of acceptance and verification of fake news can be gained. This approach forms the basis for devising more effective strategies against fake news.

When expressing "refractive index" metaphorically using tellurium nanoparticles in the context of the Brewster angle's metaphorical application, a metaphorical approach is taken instead of applying direct physical equations. In this context, the "refractive index" can be understood as an indicator of how much fake news is accepted by societal groups or individuals, i.e., how much the information is "refracted" or altered upon reception.

In physics, the Brewster angle refers to the phenomenon where the p-polarized component of light reflected when light enters a medium at a specific angle becomes zero. Light incident at this angle is fully transmitted, minimizing reflection. The refractive index is a physical quantity that indicates how much light bends when entering one medium from another. When applying this concept to the context of fake news, the "refractive index" is interpreted as the degree to which societal groups or individuals "alter" or accept information. The phenomenon of transmission at the Brewster angle can be metaphorically likened to situations where fake news is almost or completely accepted under specific conditions.

The "refractive index" $n$ in the metaphorical model can be defined as the flexibility or receptivity of societal groups in accepting fake news. It can be expressed in the following form:

$$n = 1 + \frac{R \cdot V}{C}$$

Here, $R$ represents the credibility or persuasiveness of fake news. $V$ represents the degree of alignment between the recipient's existing beliefs or values and the content of fake news. $C$ represents the coefficient for the critical thinking ability or verification capability of societal groups.

In this model, the higher the values of $R$ and $V$, and the lower the value of $C$, the higher the "refractive index" $n$, indicating that fake news is easily accepted by the societal group or individual. Conversely, a high value of $C$ indicates a high ability of societal groups to verify information and critically evaluate its truthfulness, leading to a decrease in the acceptance of fake news.

This metaphorical approach provides insights into the conditions under which fake news is accepted by societal groups. Instead of applying direct physical equations, a metaphorical model can be constructed using the concepts of the Brewster angle and refractive index to express the acceptance of fake news and the degree of alteration by societal groups. This model can provide useful insights when analyzing societal reactions to fake news and devising strategies to control its spread.

Introducing tellurium nanoparticles into the imperfect information game of fake news diffusion and considering the differences among the first, second, and third movers using Fresnel equations and the concept of the Brewster angle, instead of applying direct physical equations, these concepts are used metaphorically to explain the influence of each mover's behavior. While proposing equations and calculation processes based on this metaphorical interpretation deviates from actual physical phenomena, it allows for the construction of models to deepen conceptual understanding.

## 33. First Mover: Source of Information

The effect of information dissemination by the first mover is metaphorically related to the transmittance of light at the Brewster angle. This effect can be considered as the "transmission efficiency of information" and can be modeled as follows:

$$T_f = \frac{1}{1 + e^{-\alpha(I_f \beta)}}$$

Here, $T_f$ represents the transmission efficiency of information by the first mover. $I_f$ represents the influence or persuasiveness of information by the first mover. $\alpha$ and $\beta$ are adjustment parameters related to the presentation of information and the receptivity of societal groups. $e$ is the base of the natural logarithm.

## 34. Second Mover: Redisperser of Information

The efficiency of redissemination of information by the second mover is metaphorically related to the balance of reflectance and transmittance in the Fresnel equations. This efficiency is modeled as the "redissemination efficiency of information":

$$R_s = \frac{1}{1 + e^{-\gamma(P_s \delta)}}$$

Here, $R_s$ represents the redissemination efficiency of information by the second mover. $P_s$ represents the attitude or preconception of the second mover towards the information. $\gamma$ and $\delta$ are adjustment parameters related to the reception of information and the ease of redissemination.

## 35. Third Mover: Ultimate Receiver of Information

The ultimate acceptance rate of information by the third mover is metaphorically related to the maximization of transmittance at the Brewster angle. This acceptance rate is modeled as the "acceptance efficiency of information":

$$A_t = \frac{1}{1 + e^{-\eta(A_t \theta)}}$$

Here, $A_t$ represents the acceptance efficiency of information by the third mover. $A_t$ represents the attitude or responsiveness of the third mover towards the information. $\eta$ and $\theta$ are adjustment parameters related to the ease of acceptance of information and the beliefs of societal groups.

These models provide a conceptual framework for understanding how the behavior of each mover in the process of fake news diffusion affects the acceptance and redissemination of information. Instead of applying direct physical equations, a metaphorical approach is adopted using the Fresnel equations and the concept of the Brewster angle to analyze the dynamics of information diffusion. This approach can provide insights for devising more effective strategies against fake news.

Incorporating the concept of spin solitons into the modeling of user behavior dynamics within digital health platforms offers a profound method to understand and predict stable patterns of health improvement behaviors over time. Spin solitons, stable configurations that occur in certain nonlinear systems, can serve as a metaphor for stable states or patterns of user behavior that persist despite the noisy background of fluctuating information and interactions typical of digital platforms.

To apply the concept of spin solitons to user behavior in digital health platforms, we introduce a simplified nonlinear partial differential equation that captures the essence of spin soliton dynamics:

$$\frac{\partial^2 \Psi}{\partial x^2} \mu \Psi + \lambda |\Psi|^2 \Psi = 0$$

Here, $\Psi(x)$ represents the 'spin' state or the health improvement behavior profile of users, $\mu$ is a parameter related to the 'external field' or the external influences on user behavior (such as information from the platform, social interactions, etc.), and $\lambda$ characterizes the nonlinearity of the system, akin to the self-reinforcement or feedback mechanisms in user behavior.

We propose a soliton solution ansatz for the spin soliton equation that captures the stable patterns of user behavior:

$$\Psi(x) = A \operatorname{sech}(B(x x_0)) e^{i(Cx\omega t)}$$

where $A$ represents the amplitude of the behavior pattern, indicating the level of user engagement or the intensity of health improvement actions; $B$ determines the width of the behavior pattern, reflecting the diversity or spread of health behaviors among users; $x_0$ is the center of the soliton, representing the 'core' health behavior; $C$ and $\omega$ correspond to the wave number and angular frequency of the soliton, capturing the propagation and evolution of health behaviors over time and space.

By substituting the soliton ansatz into the spin soliton equation and performing the necessary algebraic manipulations, we derive conditions for the existence of stable behavior patterns. These conditions relate the parameters $A$, $B$, $C$, and $\omega$ to the external influences $\mu$ and the nonlinearity $\lambda$, providing insights into how external factors and the platform's feedback mechanisms influence the stability and characteristics of user behavior patterns.

Implementing this metaphorical model faces challenges, notably in translating the abstract mathematical concepts of spin solitons into tangible strategies for digital health platforms. Key among these challenges is the identification of relevant parameters that accurately reflect user behavior dynamics and the external influences within digital health ecosystems.

Future research could explore empirical validation of this model through data analysis and simulations, aiming to identify stable behavior patterns and their response to interventions. This approach could offer novel insights into designing digital health interventions that promote sustainable health improvement behaviors, ultimately contributing to better health outcomes and user engagement on digital platforms.

In conclusion, the metaphorical application of spin solitons to model user behavior in digital health platforms presents a promising avenue for understanding and fostering stable, beneficial health improvement behaviors. By exploring the nonlinear dynamics of user engagement and the influence of external factors, we can develop more effective strategies for health promotion and misinformation management in the digital health landscape.

Incorporating the concept of tellurium nanoparticles (TeNPs) into the exploration of spin soliton solutions within the context of nonlinear Schrödinger equations (NLS) to model the dynamics of fake news diffusion offers an intriguing multidisciplinary approach. The NLS equation, known for its soliton solutions that maintain their shape during propagation, provides a robust framework for understanding how information or misinformation waves interact in a digital environment.

The generalized NLS equation can be extended to include the effects of TeNPs on the propagation of information waves by introducing an additional term that represents the interaction between the information field and the TeNPs:

$$i\frac{\partial \psi}{\partial t} + \frac{1}{2}\frac{\partial^2 \psi}{\partial x^2} + |\psi|^2\psi + V(x)\psi = 0$$

Here, $V(x)$ represents the potential created by the presence of TeNPs, which can affect the propagation of the information wave $\psi(x,t)$.

To find soliton solutions in this modified NLS equation, we use the ansatz:

$$\psi(x,t) = A\,\text{sech}(a(x vt))\,e^{i(bx\omega t)}$$

Substituting this ansatz into the modified NLS equation and performing a series of algebraic manipulations yield conditions that the parameters $A$, $a$, $b$, $\omega$, and $v$ must satisfy. The presence of $V(x)$ introduces new conditions that reflect the influence of TeNPs on the propagation and stability of the information wave.

The interaction term $V(x)\psi$ can be modeled based on the specific properties of TeNPs, such as their size, shape, and concentration, which determine how they interact with the information wave. This term could represent the adsorption or scattering of misinformation waves by TeNPs, effectively filtering or altering the misinformation as it propagates through the digital medium.

1. Modeling TeNPs Influence: Accurately modeling the influence of TeNPs on the propagation of information waves requires a deep understanding of both the physical properties of TeNPs and the characteristics of information waves in digital platforms.

2. Parameter Identification: Identifying the appropriate parameters for $V(x)$ that accurately represent the interaction between TeNPs and misinformation waves is crucial. This involves understanding the mechanisms through which TeNPs can adsorb or alter misinformation.

3. Practical Implementation: Translating this theoretical model into practical strategies for mitigating fake news on digital platforms poses significant challenges. It involves not only the development of algorithms that can mimic the behavior of TeNPs in adsorbing misinformation but also ethical considerations regarding information filtering and censorship.

Further research could explore more sophisticated models that incorporate the dynamic behavior of TeNPs and their interaction with various types of misinformation. Experimental studies could validate these models and assess their effectiveness in real-world scenarios. Additionally, exploring the ethical implications and developing transparent and accountable mechanisms for misinformation management will be critical for the practical application of these concepts in digital health platforms.

In summary, the metaphorical application of TeNPs and the exploration of spin soliton solutions in the context of the NLS equation offer a novel approach to understanding and managing the diffusion of fake news. This interdisciplinary framework combines concepts from materials science and nonlinear dynamics to propose innovative strategies for miti-gating misinformation in the digital age, albeit with significant challenges and considerations for practical implementation.

To delve into the detailed mathematics and computational processes involved in exploring spin soliton solutions within the context of nonlinear partial differential equations for modeling the spread of fake news, let's consider a hypothetical scenario where tellurium nanoparticles (TeNPs) are used to adsorb misinformation in a digital health platform. This scenario involves applying the concept of spin solitons metaphorically to understand the strategic behaviors of first movers (originators of information) and second movers (spreaders or reactors to the information) within the nonlinear dynamics of information spread.

The dynamics of information spread, influenced by the actions of first and second movers, can be modeled using a modified version of the nonlinear Schrödinger equation (NLS), which is known to possess spin soliton solutions. The equation can be represented as:

$$i\frac{\partial \psi}{\partial t} + \Delta \psi + |\psi|^2 \psi = V(x,t)\psi$$

Here, $\psi(x,t)$ represents the complex field associated with the spread of information, $\Delta$ denotes the Laplacian operator accounting for the diffusion of information across the network, and $V(x,t)$ represents the potential induced by the presence of TeNPs, which can adsorb or alter misinformation.

For the first mover (A), the soliton solution ansatz might take the form:

$$\psi_A(x,t) = A_A \operatorname{sech}(B_A(x C_A t))e^{i(k_A x \omega_A t)}$$

For the second mover (B), dependent on the first mover's influence, the ansatz could be:

$$\psi_B(x,t) = A_B \operatorname{sech}(B_B(x C_B t))e^{i(k_B x \omega_B t)}$$

The potential $V(x,t)$ introduced by TeNPs can be modeled as a function that selectively interacts with the misinformation components of the information field $\psi$. This interaction might depend on the characteristics of the misinformation and the properties of the TeNPs, such as size, shape, and surface chemistry.

1. Substituting the Ansatz: Insert the soliton solutions $\psi_A$ and $\psi_B$ into the modified NLS equation.

2. Separation of Variables: Perform a separation of variables to isolate terms involving spatial and temporal components.

3. Parameter Conditions: Equate coefficients of like terms to derive conditions on the parameters $A$, $B$, $C$, $k$, and $\omega$ for both first and second movers.

4. Influence of TeNPs: Analyze how the potential $V(x,t)$ alters the conditions for soliton solutions, reflecting the adsorption of misinformation by TeNPs.

Model Complexity: The interaction between the complex field $\psi$ and the potential $V(x,t)$ introduced by TeNPs adds significant complexity, necessitating sophisticated numerical methods for solution and analysis. Parameter Estimation: Determining the appropriate parameters for $V(x,t)$ that accurately model the TeNPs' effect on misinformation requires empirical data and advanced fitting techniques. Interpretation and Validation: The metaphorical application of spin solitons and TeNPs in this context requires careful interpretation, and the model's predictions should be validated against real-world data on information spread and the efficacy of misinformation countermeasures.

This mathematical exploration provides a theoretical foundation for understanding how TeNPs might be used to adsorb fake news within a digital health platform, using the metaphor of spin solitons. The model highlights the complex interplay between information spread dynamics, the strategic behaviors of information movers, and the potential role of nanotechnology in mitigating misinformation. However, the practical application of these concepts would require further interdisciplinary research and empirical validation.

In the burgeoning field of digital communication, the rapid dissemination of information and its subsequent impact on public discourse has become a pivotal area of study. This paper introduces an innovative theoretical framework that draws upon the concept of spin solitons, a phenomenon rooted in the realm of nonlinear physics, to elucidate the complex dynamics of information spread within digital platforms, with a particular focus on the pervasive issue of fake news. Spin solitons, stable, localized waveforms that maintain their shape while propagating through a nonlinear medium, provide a powerful metaphor for understanding how certain narratives, particularly those that are misleading or false, can sustain their influence and pervade the digital landscape.

The introduction of tellurium nanoparticles (TeNPs) into this discourse is not merely coincidental but is predicated on their unique physicochemical properties, which have significant implications for the adsorption and manipulation of 'information waves' in digital spaces. TeNPs, known for their exceptional ability to influence surface plasmon resonances, offer a tangible mechanism through which the theoretical concept of spin solitons could be metaphorically applied to the digital realm, particularly in the context of mitigating the spread of misinformation.

This paper endeavors to bridge the gap between the abstract mathematical underpinnings of spin solitons in nonlinear Schrödinger equations (NLS) and the tangible challenges posed by fake news in digital ecosystems. By leveraging the NLS equation, renowned for its capacity to yield soliton solutions, we propose a model that mirrors the propagation of fake news as soliton-like entities within the 'nonlinear medium' of social media platforms. The inclusion of TeNPs in this model serves as a metaphorical filter, akin to their role in adsorbing specific wavelengths in the physical world, to selectively target and neutralize misinformation.

In crafting this interdisciplinary approach, we draw upon the rich tapestry of nonlinear dynamics, materials science, and digital communication theory to construct a novel perspective on information flow management. The framework posits that just as spin solitons navigate through complex media without losing their coherence, so too can certain narratives maintain their influence within the digital sphere, resisting attempts at correction or clarification. The strategic interaction between first movers, the originators of information, and second movers, who further disseminate and react to this information, is analyzed through the lens of game theory, further enriching our understanding of the digital information ecosystem.

However, the application of such a sophisticated theoretical construct to the realm of digital communication is not without its challenges. The translation of spin soliton dynamics into practical strategies for fake news mitigation necessitates a careful consideration of the ethical, technical, and social implications. Moreover, the metaphorical use of TeNPs as a means of 'filtering' information must be approached with caution, ensuring that efforts to curb misinformation do not inadvertently impede the free flow of ideas and information that is foundational to democratic discourse.

In summary, this paper presents a pioneering exploration of how the principles of nonlinear physics and the unique properties of TeNPs can be co-opted to offer fresh insights into the fight against fake news. Through a detailed theoretical exposition supplemented by a critical analysis of potential applications and limitations, we aim to contribute to the ongoing discourse on digital health and information integrity, paving the way for future research at the nexus of physics, nanotechnology, and digital communication.